# Rotational Symmetry and the Transformation of Innovation Systems in a Triple Helix of University-Industry-Government Relations

*Technological Forecasting & Social Change* (in press)

Inga A. Ivanova [a]* & Loet Leydesdorff [b]

**Abstract**

Using a mathematical model, we show that a Triple Helix (TH) system contains self-interaction, and therefore self-organization of innovations can be expected in waves, whereas a Double Helix (DH) remains determined by its linear constituents. (The mathematical model is fully elaborated in the Appendices.) The ensuing innovation systems can be expected to have a fractal structure: innovation systems at different scales can be considered as spanned in a Cartesian space with the dimensions of (*S*)cience, (*B*)usiness, and (*G*)overnment. A national system, for example, contains sectorial and regional systems, and is a constituent part in technological and supra-national systems of innovation. The mathematical modeling enables us to clarify the mechanisms, and provides new possibilities for the prediction. Emerging technologies can be expected to be more diversified and their life cycles will become shorter than before. In terms of policy implications, the model suggests a shift from the production of material objects to the production of innovative technologies.

---

[a] School of Business and Public administration, Far Eastern Federal University, 8, Sukhanova St., Vladivostok 690990, Russia; inga.iva@mail.ru ; * corresponding author.
[b] Amsterdam School of Communication Research (ASCoR), University of Amsterdam, Kloveniersburgwal 48, 1012 CX Amsterdam, The Netherlands; loet@leydesdorff.net .

# 1 Introduction

A market-oriented economy's transition to a knowledge-based economy increases the pressure of globalization because dynamics can be expected to change at the supra-national level. In this study, we argue that the conceptualization of this system in terms of a three-dimensional vector space as, for example, specified in the so-called Triple Helix of university-industry-government relations [1], provides the sufficient and necessary conditions for the specification of a mathematical model that can explain how technological trajectories can be formed between "double helices" (DH), and how a self-regenerating system can be expected to develop at the global level of a Triple Helix (TH). We illustrate how the communication field generated by the interactions among the trajectories is sensitive to the order of the relations. Thus, (linear) symmetry is broken and innovation can be expected to emerge.

A system's approach to innovation studies was first introduced by Freeman [2] with reference to the Japanese system of innovations. The approach was then generalized by Lundvall [3, 4] and Nelson [5] to the theory of "national systems of innovation." Porter [6, 7] abstracted from the national context by focusing on "clusters" of innovations that can be more dense and differently shaped in regional and/or national settings. Gibbons *et al*. [8] added that "the new production of scientific knowledge" transforms the systems dynamics from "Mode-1" into a trans-national and trans-disciplinary field that is driven by communication across institutional borders ("Mode-2").

Leydesdorff [9] specified that a system with three subdynamics can endogenously generate complex dynamics, but in the Triple Helix metaphor [1, 10] the emphasis initially remained on integration in terms of institutional relations. Leydesdorff [11, 12] then distinguished between this neo-institutional model of relations, and the neo-evolutionary model of different subdynamics such as wealth generation, novelty productions, and normative control in any system of innovations. These subdynamics can also be considered as functions and then be modeled as vectors in a vector space.

The paradox of the current situation is that "if the working of the Triple Helix (…) is relatively well explored and usually examined at a specific moment in time (a synchronic interaction), a methodology for analyzing the transition among Triple Helix regimes over time (a diachronic interaction) is a relatively under-conceptualized problem" [13, p.2]. Hitherto, the TH model has rested mostly on phenomenological case-studies. The failure to understand the mechanisms causing the dynamic evolution of the Triple Helix significantly reduces the effectiveness of this model. Case studies describe situations in different regions and are difficult to compare. In our opinion, one should avoid thinking in terms of phenomenological descriptions and instead develop analytical techniques that enable us to study how different factors interact in a systemic context. In this study, we claim that such a TH model can be specified on the basis of formal logic, and then elaborated into a mathematical formulation.

The research question of this paper is to overcome the drawbacks of the phenomenological approach by presenting a mathematical formulation of the TH model. This can help to formalize and operationalize the non-linear dynamics and reveal the features that remain hidden in phenomenological descriptions. How does the interaction among the three players—Industry (or Business), University (Science), and Government—develop an innovation infrastructure? From this perspective, the TH model is special not only because it allows us to create an effective system for the development and promotion of innovations, but also because it provides the lens through which one can make a breakthrough in understanding the fundamental mechanisms in innovation systems. The non-linear dynamics of interaction among actors can be expected to lead to a fractal structure in a TH system. Such a fractal structure provides *self-similar* patterns in innovation systems at different scales, which are replicated in innovation activities at various scale levels. Because of this fractal structure [14], innovations can be integrated into systems not only nationally, regionally, or sectorially, but across dimensions while incorporating both separate companies and projects. At each scale a TH structure can be expected and further analyzed.

This paper is organized as follows. In Section 2, the classical TH model is described in terms of innovation systems. In Section 3, we discuss the issue of evolutionary symmetry in a TH system. The TH model can be described mathematically as a group of rotational symmetries in a three-dimensional space. In Section 4, we focus on innovation cycles and waves. The cyclical character of innovations, that is, the periodic arising of innovative activity, allows for defining waves of innovations and describing innovation as propagating in a specifically defined space. In Section 5, the combination of TH symmetry with innovation waves is shown to result in non-linearity and self-organization. In Section 6, we explain how the interactions among different technological trajectories can be expected to result in the fractal structure of the TH model. In Section 7, the results are summarized and we elaborate on options for policy-makers. Readers especially interested in the mathematics will find a more elaborate description of the model in the Appendices.

## 2  The Triple Helix model of Innovations

The Triple Helix model assumes that the driving force of economic development in the post-industrial stage is no longer manufacturing, but the production and dissemination of socially organized knowledge. Institutions that generate knowledge increasingly play a role in the networks of relations among the key actors: University (*Science*), Industry (*Business*), and Government (*Governance*). The spheres of these activities are increasingly overlapping. In areas of intersection, the actors can partially substitute for one another.

Universities, for example, in addition to fulfilling educational and research functions, increasingly undertake a part of the business functions, creating small innovative companies and becoming thus a stakeholder in socio-economic development. Industrial corporations create their own research centers and training centers for employees. They can also use the university's infrastructure in order to conduct their own R&D activities, and thus shift part of their costs to the state as the main source of funding for universities. Governments encourage the development of small innovative enterprises through priority financing of specific universities and legislative regulation, and they stimulate industry to develop and implement new innovative technologies. Universities and industry can partially substitute for the state in the creation of an innovation infrastructure. The overlapping institutional spheres of these three actors are graphically represented in Fig. 1.

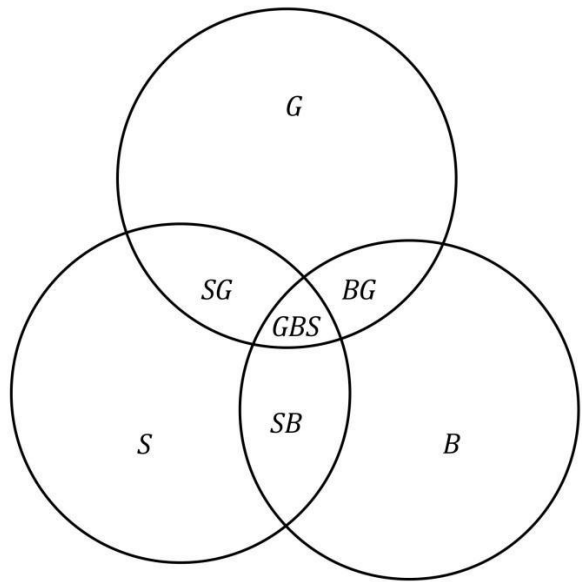

**Figure 1**: A balanced Triple Helix model; *S*: science; *G*: government; and *B:* business.

The domain of the TH model coincides with the area where the institutional spheres of the three actors—*S, B,* and *G*—overlap, and where there is maximum interaction among these actors. However, the respective area sizes and the nature of the agents interrelating can be expected to change constantly due to the interactions. Constant change is therefore one of the TH model's features; the other feature is the model's non-linearity.

Despite the wide acceptance of the Triple Helix model, there remain a number of issues requiring further attention. For example, one can ask: which mechanisms are responsible for the increased potential for coalition building and how can one understand the non-linearity of the model? From a mathematical point of view, this "non-linearity" suggests a non-linearity in the functional dependencies. What kind of functional dependency is meant in the case of a TH model? Can it be described as an exponential function or is it a power function, a sinusoidal or something else? Without a mathematical model, one cannot detect the functional dependency; and without a proper definition of the nonlinearity involved, one cannot establish a link between the interaction of actors and the innovations generated at the systems level.

The nonlinearity of modern innovation systems is due to the nonlinear nature of the economic development in such systems. Freeman [2] pioneered the vision that innovation should be understood as an interactive process, and not as a linear one. This nonlinear model deviates from Schumpeter's [15] model, because he distinguishes between creating innovations and the process of their development and application, and accordingly suggests that the number of innovations would be directly proportional to the number of inventions—providing a linear function. However, the current paradigm of innovation systems is nonlinear.

This nonlinearity is caused by the reverse process of transferring information from the subsequent stages of advancement to previous ones, in addition to the direct process of technology transfer from R&D to the market. The market is not only driven by innovations, but, in turn, acts as a stimulant of innovations. Thus, the number of innovations will nonlinearly depend on the number of inventions. However, the nonlinear process of technology transfer is also typical for systems in which there is only a dyadic interaction present, that is, Double Helix (DH) systems, but despite the nonlinear process these systems, as we show below, cannot be considered as *non-linear systems*.

How is the non-linearity of innovation processes in systems with dyadic interaction different from nonlinearity of the processes in systems with a triadic interaction? Non-linearity at the systems level can be defined as the ability of self-replication and self-generation of new organizational formats. This non-linearity is based on phenomena that can be witnessed in the area of the triple intersection, where the actors can mutually replace each other, in addition to being the center of generating new innovative technologies. The intersection can thus become the center of new organizational formats.

We shall show (in Section 5) that such a nonlinear interaction is only possible if the number of interacting actors in the same field is larger than or equal to three. In this sense, a model (as in Fig.2), in which only dual intersecting areas are represented and the triple intersecting area is missing, cannot be considered nonlinear, and will not be able to generate new organizational formats, even though all types of bilateral interactions (*SG, SB,* and *BG*) are present.

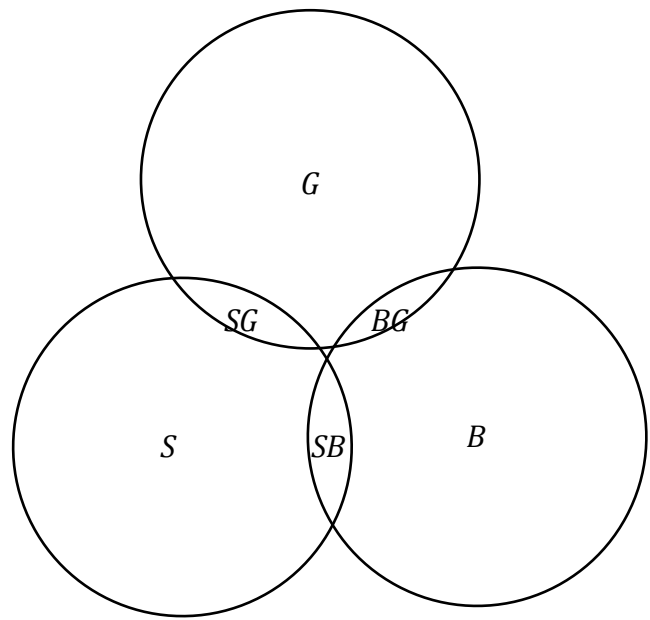

**Figure 2:** An example of a model without a triple intersecting area.

The continuous variation and nonlinearity makes the Triple Helix a nonlinear dynamical system. A nonlinear dynamical system must have the following features: first, the system contains feedback loops; second, areas are present where more than a single state of equilibrium is possible; third, the system can be considered as fractal; and fourth, there is a sensitive dependence of the systems dynamics on initial conditions [16]. All these features must be appropriately accounted for in a mathematical model of the Triple Helix.

## 3  The Triple Helix symmetry and evolution

In this section, we describe the symmetry aspects of the TH model that arise from its evolutionary development. A TH system can be conceptualized in terms of its components, relationships, and functions [13]. Components include the institutional spheres of Business ($B$), University ($S$), and Government ($G$); relationships reflect links among these actors; and functions reflect the results of their activities. If one examines the relative influence of institutional spheres, one can distinguish three TH regimes: TH I - the state plays a dominant role, where government drives science and industry; TH II - the market plays a dominant role (a so-called laissez-fair regime); and TH III – a knowledge-based economy in which the role of science increases greatly [1]. Leydesdorff and Meyer [17] suggested presenting this model in the three dimensions of a Cartesian space, with axes respectively corresponding to the three *functions* of Wealth generation ($W$), Novelty production ($N$) and Legislative control ($L$). Fig. 3 shows patents as exemplary events in this three-dimensional space of Triple Helix interactions.

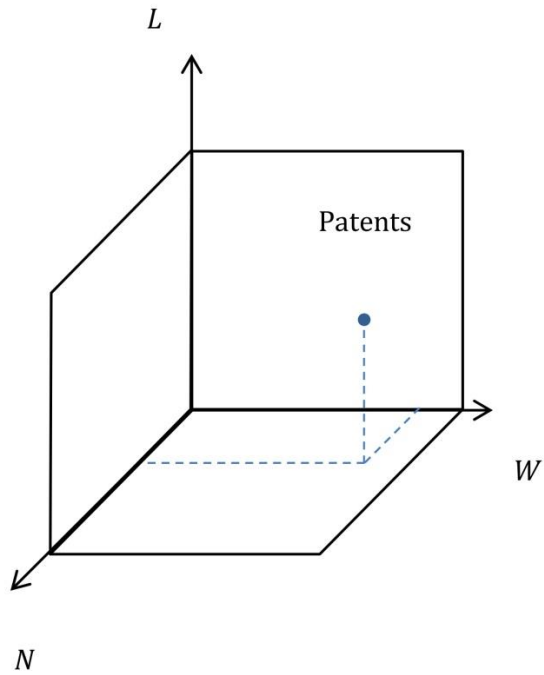

**Figure 3:** Patents as events in the three-dimensional space of Triple Helix interactions. (Source: [18]).

The events in Fig. 3 can also be presented as a vector *P*, drawn from the origin of the coordinate system to the point marked by the word "patents." The input of *W, N, L* to patents can be represented as their corresponding shares per unit patent, so that the vector *P* is normalized. We can then decompose vector *P* into the sum of three vectors: $P_W$, $P_N$, $P_L$, positioned along the corresponding coordinate axes *W, N, L*:

$$\vec{P} = \vec{P}_W + \vec{P}_N + \vec{P}_L \tag{1}$$

Patents can be considered as results of scientific research. Expression (1) shows that scientific research (or Science as an institutional sphere with the function of Novelty Production; Wealth generation; and Legislative control) interact in the events. Note that, on the one side, Science contributes to Wealth generation, Novelty production, and Legislative control, but, on the other side, Wealth generation, Novelty production, and Legislative control contribute to Science. To this end, expression (1) can also be written as:

$$S = b_1 W + b_2 N + b_3 L \qquad (2)$$

Coefficients $b_1, b_2, b_3$ reflect in this case the relative contributions of Wealth generation, Novelty production, and Legislative control to scientific research.

Analogously, one can map the results of Government and Business activity in terms of the three functions. Each of the three actors participates in Wealth generation, Novelty production, and Normative control. From a formal viewpoint, this can be expressed in the following way:

$$W = a_{11} G + a_{12} S + a_{13} B$$
$$N = a_{21} G + a_{22} S + a_{23} B \qquad (3)$$
$$L = a_{31} G + a_{32} S + a_{33} B$$

There is no unit scale for the three axes. The nature, structure, and motivation of the functions are highly different: Novelty can be measured, for example, in terms of patent citations or technologies in which these patents are used, Wealth can be measured in terms of profit, and Normative control is measured in terms of legislative acts. In sum, we map these units relatively and without dimension. The coefficients in Eq. (3) reflect the relative role of each actor in the corresponding function.

Expression (3) can be considered as a transformation from the coordinate system (*G, S, B*) to a new coordinate system (*W, N, L*). It also can be reversed and one can write:

$$G = b_{11}W + b_{12}N + b_{13}L$$
$$S = b_{21}W + b_{22}N + b_{23}L \qquad (4)$$
$$B = b_{31}W + b_{32}N + b_{33}L$$

This set of equations describes the transformation of coordinate system (*W, N, L*) into coordinate system (*G, S, B*). The coefficients in Eq. (4) can be interpreted as relative inputs of the corresponding institutional spheres to Wealth generation, Novelty production and Normative control. Expressions (3) and (4) show that TH system can be regarded from a unified viewpoint. System functions and components are interdependent.

The schema of Fig. 1 can now conveniently be represented in the Cartesian coordinate system of Fig. 4. A Triple Helix system is thus represented as a unity vector *V*. Components of vector *V* along the coordinate axes *S, B,* and *G,* correspond to the relative roles played by the institutional spheres of University, Industry, and Government. Each agent's relative role can be defined in terms of relative partaking in Novelty and Wealth production, and Normative control insofaras generated by this actor.

One can assume that the three functions are of equal importance *ex ante*. Or the relative significance of each function can be accounted by inserting the corresponding weight coefficients to distinguish among the relative importance of Wealth, Knowledge and Normative control. For example, Government can be responsible for generating 5% of Wealth, 10% of Novelty and 85% of Normative control acts ($b_{11} = 0.05;\ b_{12} = 0.1;\ b_{13} = 0.85$); University for generating 15% of Wealth, 60% of Novelty and 5% of Normative control acts ($b_{21} = 0.15;\ b_{22} = 0.6;\ b_{23} = 0.05$); and Industry can be responsible for generating 80% of Wealth, 30% of Novelty and 10% of Normative control acts ($b_{31} = 0.8;\ b_{32} = 0.3;\ b_{33} = 0.1$).

In this case, Government's contribution to the functional results of the TH can be presented as the projection of vector $V$ on the axis $G$ or: $V'_G = \sqrt{b_{11}^2 + b_{12}^2 + b_{13}^2} = 0.81$; University's contribution: $V'_S = \sqrt{b_{21}^2 + b_{22}^2 + b_{23}^2} = 0.62$; and Industry's contribution: $V'_B = \sqrt{b_{31}^2 + b_{32}^2 + b_{33}^2} = 0.86$. We can further normalize vector $V$ setting: $V^2 = 1$, then the corresponding components of vector $V$ are specified as follows: $V_G = V'_G/V^2$; $V_S = V'_S/V^2$; $V_B = V'_B/V^2$. Note that the measure units of all three axes are identical and dimensionless. The regimes of TH I, TH II, TH III discussed above correspond to cases in which the vector $V$ is directed along the $G$, $B$, and $S$ axes, respectively. Cases when the vector $V$ is positioned in the coordinate planes *SG, BG, SB,* indicate the various Double Helix models.

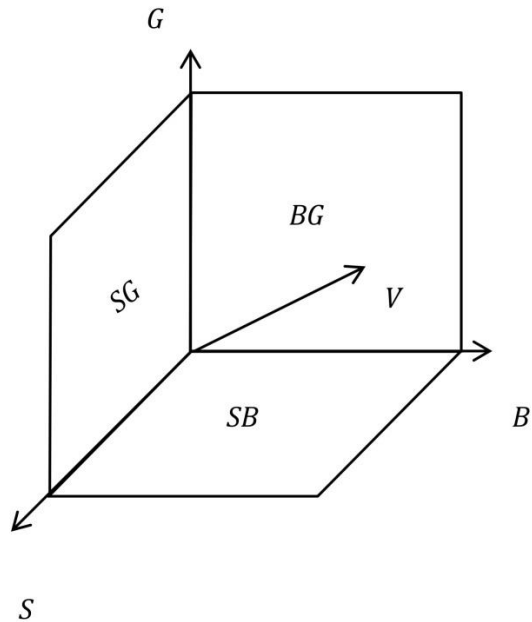

**Figure 4**: Cartesian coordinate representation of the TH model.

Over time, the relative inputs of institutional spheres in Wealth generation, Novelty production, and Normative control are subject to change. This means that the coefficients in Eqs. (3) and (4) change over time and this can be accounted for by a rotation of vector *V* in the coordinate space of (*G, S, B*).

Etzkowitz and Ranga [13] suggested describing the process of interactions and evolution via Knowledge, Consensus, and Innovations Spaces. These spaces are related to the three functions, respectively: novelty production, normative control, and wealth generation [17]. Each of the Spaces involves activity of all three actors, but the weights of the actors in each of the Spaces are unequal. The Knowledge Space is based on R&D; R&D activities creating new knowledge are mostly conducted in universities: the University can be expected to play the prime role in the Knowledge Space. The Consensus Space is primarily defined by governance with a prevailing role for Government, and the Innovation Space is based on knowledge-based entrepreneurship, encompassed mostly by the Industry sphere.

Depending on specific initial conditions in various regions, the innovation process may comprise consecutive initiatives that lead to building the mentioned spaces in various time successions. Etzkowitz & Ranga [13, 19] discuss the situation in two regions: Norrköping in Sweden and New England in the United States. While in the first region the sequence of space generation was Consensus → Innovation → Knowledge Space, in the second it was Knowledge → Consensus → Innovation Space. The creation of Spaces, however, entails a change of the corresponding actors' relative roles.

For example, the Consensus → Innovation → Knowledge Space sequence in Sweden reflects a shift of emphasis from Government to Industry, and then to Science. This process can schematically be depicted as a rotation of the vector *V* in Fig. 4 in the three dimensions of the Cartesian space. The rotation changes the relative value of the vector components, and appropriately the corresponding contributions of *G, B, S,* as institutional spheres. In other words, the evolution of the Triple Helix initiated by interactions among institutional spheres causes rotations of the vector *V*, which represents the evolution of the TH system. Because of the continuous evolution of the system, the TH symmetry can also be considered as a dynamic symmetry.

In a system with three coordinates, as against one with two, the order of the successive rotations is important. Figs. 5 and 6 show this. If one first rotates a vector *V* in the rectangular area counter-clockwise at $90^0$ around the *x*-axis and then counter-clockwise around the *y*-axis, the result is as pictured in Fig. 5c. If one first rotates counter-clockwise at $90^0$ around *y*-axis and then counter-clockwise around *x*-axis the result is as pictured in Fig. 6c. In other words two successive rotations cannot be interchanged without changing the outcome.

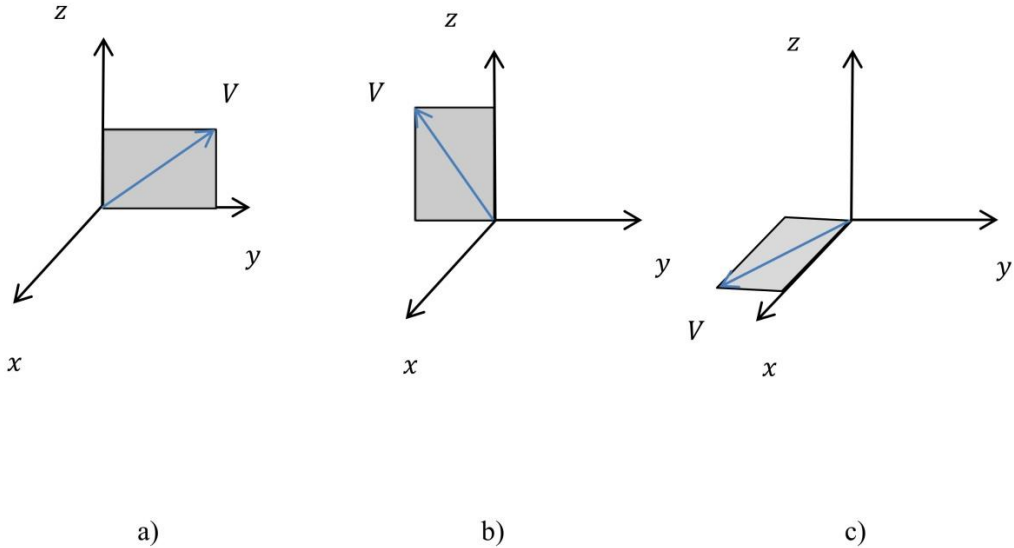

a)　　　　　　　　　　　b)　　　　　　　　　　　c)

**Figure 5**: Illustration of two successive rotations of vector *V* in a three-dimensional Cartesian coordinate system.

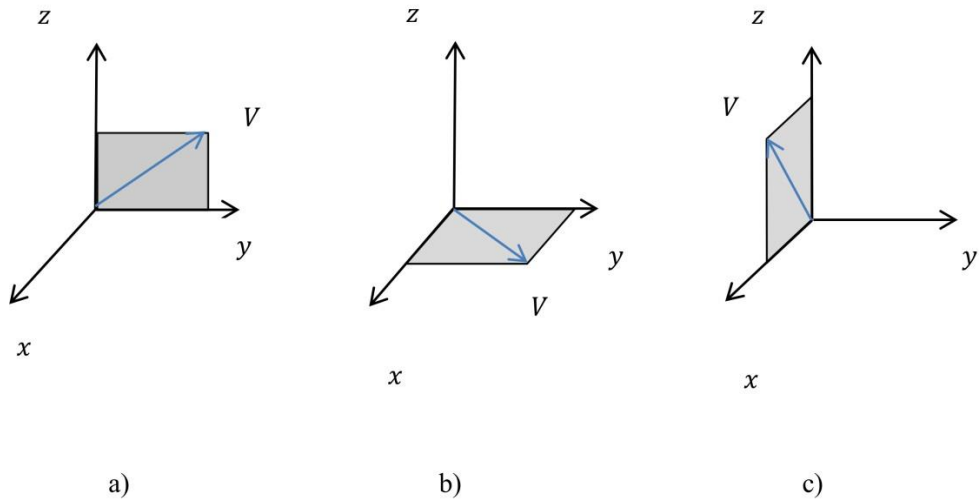

**Figure 6**: The same as in Fig. 5, but the order of successive rotations is changed.

When the above example is related to innovation practices, the order of successive initiatives is important. If one first creates a Knowledge space and then a Consensus space, one first strengthens the role of University, and then the role of Government. This can be presented in Fig. 4 as two successive rotations of vector *V*. However, if one first creates the Consensus space and follows with the creation of Knowledge space, performed by first strengthening the role of Government and then the role of University, the final results will be different.

One can make a similar argument with respect to the Double Helix model, where there is only a dyadic interaction between two actors; this leads to the same representation as in Fig. 4. But here, the rotation of the vector *V* is performed in the coordinate system of a plane, and the order of the two successive rotations makes no difference. We shall see below that this three-dimensional rotational symmetry of the TH system is one of the reasons for its non-linearity and self-organization.

**4 Innovation cycles and innovation waves**

In addition to the distinction between institutional and functional spaces between which rotation is possible, we must introduce the difference between cycles and waves [20]. In general, cyclical processes presume that some perturbation arises periodically with time intervals, whereas waves imply that the perturbations are moving in a space. We shall argue that cycles provide the spaces for the waves which can then be considered as the up-hill and down-hill trajectories on a landscape that is evolving cyclically in terms of regimes.

Cycles have extensively been studied in economics. One distinguishes between long economic cycles with a characteristic period of 45-60 years [21], Kuznets swings that can be considered as medium-range economic waves with a period of 15–25 years [22], Juglar cycles with a period 7-11 years [23], and short (Kitchin) business cycles with a period of 3-4 years [24]. The present catalogue of historical cycles lists about two thousand economic cycles lasting from 20 hours to 700 years.

Schumpeter [15] was the first to study the relations between economic cycles and waves of innovation. According to Schumpeter, the driving force behind long-term economic fluctuations is the wavy dynamics of technical and technological innovations. This concept was further developed by Mensch [25], Dickson[26], Freeman[2], Modelsky [27, 28], Thompson [29], and Papenhausen [30]. Freeman and Perez [31] considered techno-economic paradigms as the main factor in the emergence of long Kondratiev waves.

In addition to techno-economic paradigms, one distinguishes also regulatory paradigms. Technological paradigms refer to the scientific and manufacturing base in specific fields of technology [32, 33], whereas a regulatory paradigm operates as an institutional framework. Freeman and Perez [31] entertain a dialectical or evolutionary model in which the forces of production generate new technological possibilities that periodically upset the regulatory frameworks [34].

A technological development within the limits of a specific paradigm that is molded by market regulations can be considered as a technological trajectory. Like technological trajectories, formed by markets and technologies, markets and political-decision making processes can also form similar trajectories[35], which we may call regulatory trajectories. Regulatory trajectories combine decision making and market forces; for example, into bureaucracies.

Whereas regulation is focused on the support of manufacturing in industrial economics, in a knowledge-driven economy regulatory mechanisms focus on supporting knowledge generation structures. From a formal perspective, regulatory trajectories evolve like technological ones: both market forces and decision making can be locked into technologies [36]. Regulatory trajectories are shaped in co-evolutions, just as technological trajectories at the interfaces between technology push and demand pull.

There are periods when knowledge development plays the primary role, and periods when demand plays the primary role in shaping a technological paradigm [37]. Markets and scientific R&D activities can also show co-evolutionary dynamics; for example, when a large part of university R&D is performed in consideration of present or future market demand. One may wish to call this a knowledge trajectory. As Leydesdorff and Etzkowitz [38] formulated, "three dynamics can be distinguished: the economic dynamics of the market, the internal dynamics of knowledge production, and governance of the interfaces at different levels."

Due to the cyclic nature of innovations, product technologies and associated industry structures also tend to have a life cycle [39, 40, 41]. The process of innovation development and introduction is not uniform and is characterized by periodic ups and downs, defined by economic conditions. Most active innovation is carried out in periods of economic boom, in bullish areas of economic cycles [42]. Many industries show the dynamics formed by a succession of different technologies, with one technology being dominant in one period of time and then succeeded by another technology that is dominant for another period of time, which in turn is succeeded by other technologies.

The innovation cycle is accompanied by successive activation along the chain of innovation developers, manufactures, distributors, technology adopters, regulatory decision-making officials, and consumers. This activation proceeds not all of a sudden, but step by step. "An innovation can be conceptualized as a trajectory of an idea or concept within science as an intellectual and social organization as well as within the domain of legal encoding (patents) and marketing (industry)" [43]. This process is more reminiscent of a wave than a cyclic process. In the TH model, this implies that one has to model the diachronic co-evolution of the spheres around each of the actors involved.

In summary, the innovation process can be described not as a cyclic, but as a wave process. The perturbations can be considered as innovative activities. The latter can be described by the function $c(N,t)$. The number of process participants is $N$—an integer value. In the case of a very large number of participants, we can replace the discrete variable $N$ by a continuous variable $x$. Innovation activity propagation in the space of innovation adopters can be described by a wave. Cyclic bursts of innovation activities generate a chain of waves, or a chain of successively replaced technologies. In the simplest case, these waves are a solution of the linear wave equation:

$$c_{tt} - a^2 c_{xx} = 0 \qquad (5)$$

The solution of Eq. (5) is a function of an arbitrary type: $c(x,t) = f(x \pm at)$. This function describes the spread of innovations along the longitudinal trajectory. Meanwhile, an innovation system can be expected to continuously generate new innovations of a similar type that can replace old ones. To describe the spread of this set of innovations in a uniform way, Eq. (5) can be considered an equation for not one, but many innovation waves. This multitude of waves, or a "field" of innovation waves, can be considered as an evolving technological trend, selected by the market and stabilized along a technological trajectory.

The wave equation (5) can now be interpreted as an equation for the innovation field, which describes a number of technologies comprising a technological trend. In a TH system diffusion of innovation affects all the institutional spheres—scientific, business, and regulatory. For example, achievements in the microprocessor sphere entail appearances of new business manufacturing trade and models (e.g. electronic commerce) and new legislative regulation (e.g. an electronic signature is now legislatively considered equivalent to an ink signature). Hence, the function of innovation activity $c(x,t)$ should reflect the coevolution among the three actors (*S, B,* and *G*).

Instead of one Eq. (5) we should write three equations—one for each of the functions: $c_G(x,t)$, $c_B(x,t)$, $c_S(x,t)$, reflecting the evolution of corresponding actors. This can also be achieved if we present the function $c(x,t)$ as a vector function as follows:

$$c(x,t) = \begin{pmatrix} c_G(x,t) \\ c_B(x,t) \\ c_S(x,t) \end{pmatrix} \quad (6)$$

This function can be conveniently mapped in a three-dimensional Cartesian coordinate system as in Fig. 4. The three components of this function reflect regulatory, technological and knowledge trajectories, that is, they correspond to the relative influence of *G, B,* and *S* actors over time. The internal symmetry of the function $c(x,t)$ is described by a group of rotations in the three-dimensional space *O(3)*.

In summary: we argue that technological developments, resulting from interactions among political, economic, and technological factors, take the form of waves, propagating in the space of innovation adopters. In the next sections we shall show how the combination of innovation waves with the internal symmetry leads to the nonlinearity and self-organization of a TH system.

**5  Self-organization in a Triple Helix**

The functioning of a TH system differs fundamentally from the functioning of a DH system. Whereas in a DH of university-industry relations two subdynamics shape each other in a coevolution that may lead to relatively stable trajectories, the addition of a third subdynamic, as in a TH system of university-industry-government relations, makes these trajectories unstable [44]. The TH dynamics can transform existing trajectories into new ones. In interactions among TH actors the partners act as selection environments for one another; the interactions among selection environments can generate new selection environments that then support new trajectories. This is suggestive of self-organization. But what is the mechanism of this self-organization?

In Fig. 4 above, we mapped the innovation system using three orthogonal coordinates ($G, S, B$). When innovation in one sphere occurs it can be expected to lead to a redistribution of functions among actors, and the position of the vector $V$ is changed. The rotation corresponds to the dynamic change of all agents' relative contributions due to the interaction among them. We proposed considering this change in position as a rotation of the coordinate system, but this rotation is not performed simultaneously along the whole path of an innovation trajectory. In periods when the market plays the major role in the development of the technological paradigm, the contribution of business (*B*) would be larger than that of academia (*S*). When the situation changes and technological knowledge starts to play a primary role in shaping a technological trajectory, the relative contributions of actors *S* and *B* can also be expected to change. This implies some reorganization in technological, manufacturing, communicative, and market distribution networks.

The reorganization successively affects a number of participants and cannot be carried out without delays. In the initial stages, innovation affects only the groups of participants dealing with the initial phase of innovation. This entails a redistribution of functions among actors, concerning only these groups. Other groups of innovation participants, which join the innovation process at later stages, still remain unaffected by the new redistribution of functions. When the rotation is performed step by step, change gradually diffuses along the chain of innovation participants. These two processes can be considered as global and local transformations of the TH system.

Eq. (5) describes the propagation of innovation activities along the chain of participants as a wave. This equation of motion should be written in an invariant form, i.e. it should keep its form unchanged when the system undergoes local transformations. The mathematically interested reader can find more details in Appendix A. This condition entails the appearance of additional terms to Eq. (5). The additional term may be treated as an additional compensation field coupled to the field $c(x,t)$ that describes innovation waves. Appendix B shows that in the case of the DH system equation, this additional field is linear, whereas in the case of the TH system, this additional field leads to a non-linear equation. The origin of these differences lies in the differences between DH and TH symmetries. Both systems possess rotational symmetries. But while two successive rotations in the DH system can be interchanged without changing the final result, in a TH system the final result strongly depends on the order of the rotations.

This additional compensation at the systems level can be interpreted as describing a dynamic communicative interaction. Etzkowitz and Leydesdorff [1] proposed to consider this as an additional overlay of communications in relation to university, industry, and government, shaping a potential hyper-cycle of meaning exchange processes that can operate as feedback or feedforward on the three underlying spheres of communication, and thus shape new opportunities. This communication among actors may lead to the emergence of new technologies, causing the shift of the technological trajectory. From these shifts, new communicative waves can arise that affect other technological trajectories.

The DH model refers mostly to a market or manufacturing economy. Technology trajectories, supported by large corporations, are then quite stable and subject to amendments, but less to radical change [45]. Because radical change and the creation of a new technological trajectory requires a restructuring of the existing manufacturing and marketing networks, which result in a potential reduction of marginal returns during the time of restructuring, radical innovations are not very welcome among large manufacturing corporations. According to economics, the behavior of manufacturers is defined by demand, and demand is expected to exert its influence within the limits of a technological paradigm. Hence, dyadic interaction leads to relatively stable technological trajectories, sometimes locked-in into a sub-optimal trajectory [35, 46]. The interaction in this DH model is linear, in the sense that new types of links and structures creating new organizational mechanisms are not likely.

The smallest possible number of actors for a non-linear innovation system is three, as in the Triple Helix model. In this case the region of overlapping institutional spheres is a center for the generation of new organizational formats. Let us elaborate mathematically on the prediction of the kinds of interactions among innovation and communication fields in the case of Double Helix and Triple Helix models. Fig. 7 and 8 show the respective dynamics graphically:

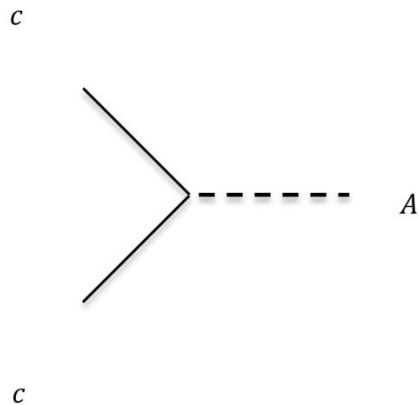

**Figure 7**: Triple vertex showing interaction between innovation $c$ and a communication field $A$ in a Double Helix system.

For a Double Helix system there can be only the kind of vertex as shown in Fig. 7, but for a Triple Helix system there can also be kinds of vertices, as shown in Fig. 8. The first describes the interaction between the innovations $c$ and the communication field, and in the second we show the additional option of self-interaction within the interaction at the level of the communication field $W$.

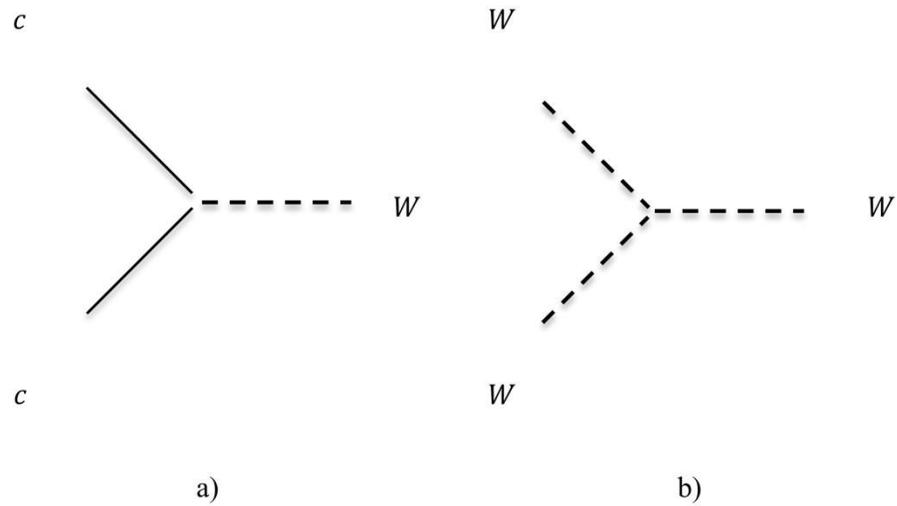

**Figure 8**: Triple vertices showing interaction between innovations *c* and communication fields *W* for a Triple Helix system.

This self-interaction arises due to the third-order non-linearity in equation (B.14) for a communication field. The interpretation is that in a Double Helix system, the communication field is only the (resulting) product of the interactions among innovations in a field. But in the case of a Triple Helix system, the communication field *W* can also function as a source itself. Communication among actors in a TH model can generate new types of links and structures due to the interactions among the institutional spheres of *G, B,* and *S*. The relationships among the agents in a TH can function as an evolving coordination mechanism that shapes and modifies institutional structures.

Whereas the expectation for bi-lateral interactions tends toward equilibrium—for example, a technological trajectory that is stable over-time, or a "lock-in"—tri-lateral interactions can de-stabilize existing trajectories and result in the transition to a new technological trajectory and, if radical enough, even to a new paradigm [35]. A new technological paradigm requires the construction of a new infrastructure and institutional modification, because a new innovation environment is prevailing. New innovation environments emerge in the TH model because of fluctuations in technological trajectories that introduce instability and in turn generate new innovation options, resulting in the ramification of an initial TH system. The resulting dynamics can be attributed to the different dynamic symmetries in the Double and Triple Helix models.

It should be noted that the TH model is unique in the sense that it possesses the minimal number of participants (actors) that provide a system with these non-linear dynamics. Possible next-order systems, i.e., prospective Quadruple Helix, Quintuple Helices, and so on [47, 48], are also expected to inherit this same non-linear feature, given the non-Abelian symmetry groups specified in the Appendices.

## 6  The fractal structure of innovation systems

Field $W$ in Fig. 8a can be considered the interaction field between two different innovation fields $c$. If the field $c$ describes the evolution of technological trends, the communication field $W$ in Fig. 8b can be considered as a field describing information exchanges between technological trends. Different technologies can influence each other; for example, the development of microelectronics and integrated circuit technology made it possible to develop the computer industry. On its turn, progress in the computer industry has allowed for computerization and improvement of integrated circuits technology. Continuous improvements occurring in the two areas and the interchange of technology can lead to the faster development of both technological trajectories because of integrating effects at the systems level [49]. The study of the interactions between technological trends, i.e., technological trend dynamics, will be the subject of our future research.

In terms of dynamic symmetry, the interaction between the various technological trends is a function of the environment in which innovation is carried out. An innovation system creates such environments, by acting as an organizer. Depending on the environment, this interaction can be both linear and nonlinear. Linear interaction suggests that technological trajectories are mutually enriched by the exchanges but will remain within their borders. In that case, the interaction between the trajectories can be expected to result in a change in the nature of the evolution of each of them without necessarily generating new trajectories. Nonlinear interaction between different trajectories, however, can lead to the formation of new trajectories and new markets. "In a complex dynamics, the three subdynamics operate upon one another, and thus upset previously constructed quasi-equilibria, leading the system into new regimes" [50].

Self-generation of the communication field *W* (as in Fig. 8b), can be recursive and then can create a tree of "virtual technologies," or expectations about other possible options, which, for example, can be represented by patents awaiting market adoption. Only a small number of these patents will result in market products. But the presence and quantity of "virtual technologies" strongly affects competitiveness in the market. An example of nonlinear interactions in the TH model is provided by the interaction between computer technologies (PC market) and communication technologies (Internet), which has led to the formation of new markets: e-commerce, social networking sites, etc.

In such cases, functional communication is codified in institutional settings forming new communicative network and overlays. These new markets are supported by relevant infrastructure elements—software developers (actor University *S*), providers of goods, services, software (actor Industry *B*), and laws, governing the relations between the participants in each sector (actor Government *G*). In turn, the development of markets for e-commerce and social networks and their interaction with the computer market resulted in the emergence and growth of the Tablet PCs and smartphones market. This new market also leads to the formation of a TH structure. Thus, more and more markets can emerge. Each market encompasses a number of technologies, and each of these technologies requires a support system with a TH structure. In this way, we expect continuous TH system cloning, resembling the formation of a fractal structure.

This brings us back to the conceptualization of the Triple Helix system. Originally, the TH model was mainly applied to national innovation systems. National innovation systems contain a set of interrelated organizations engaged in the production and commercialization of scientific knowledge and technologies within national borders. In a next stage, the concept of applying a national innovation system to smaller-scale levels: regional, sectorial systems of innovations [51], technological systems [52] and corporate innovation systems as different perspectives were proposed [53]. Furthermore, the concept can be expanded to the larger-scale level of global regions.

A knowledge-based economy can be more globally than nationally integrated, as in the case of Norway [50]. This was theoretically conceptualized as the necessity to account for additional dimensions. These authors formulated as follows: "in addition to local integration in university-industry-government relations, one should also account for the dimension globalization-localization or, alternatively formulated, the international dimension as important to further development at the national system level." In these new dimensions under the actor $G$ one should mean not national government, but the actor that more generally performs normative control functions in the systems under study.

In a previous study [54], one of us suggested that the fractal structure of the innovation system should follow from the fractal structure of economic and innovation cycles, so that a single innovation system may serve a single innovation cycle or even a single project. It was shown that this TH model can also be applied to the single project of high-tech wooden house construction. Intuitively it is clear, that a single innovation system, by virtue of its functional limitations, cannot cover the complete variety of innovation cycles, each of which also have a different nature. Ideally, each cycle of innovations must meet the appropriate scope of its innovation system. In this sense, a set of innovative systems of different scales, corresponding to a fractal set of innovation cycles, is organized like a fractal manifold.

Our argument is that TH can be expected to exhibit the characteristic of a self-replicating system. The self-replication is an intrinsic feature of the fractal manifold. And like other fractal manifold, this self-replication is performed both upwards and downwards at different scale levels.

Hitherto, it was unclear what precisely the underlying mechanism of this fractal structure formation was. Building on the conclusions above, we are able to visualize this mechanism. This fractal structure is generated as a result of nonlinear interactions in the innovation system with a number of actors of no less than three, as a consequence of local transformations in the internal symmetry of the system. One can represent these interactions in the form of pictures. Fig. 9 shows a few initial steps of the self-generation from communication fields on the basis of Fig. 8b, in which each straight line is successively and recursively substituted by a vertex.

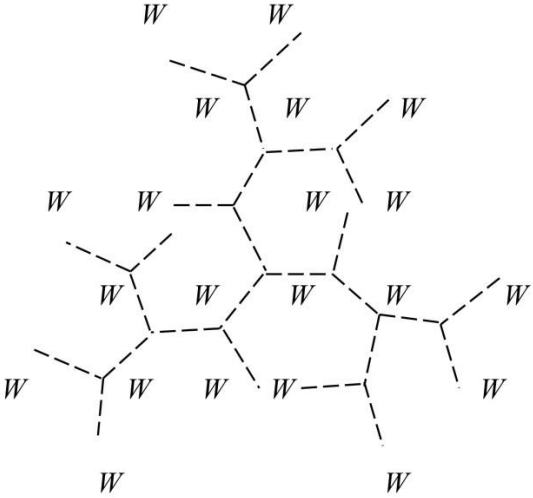

**Figure 9**: A fractal formed by the self-generation of communication field.

Each vertex can be considered as a set of virtual technologies supported by virtual organization structures that are clones of the original TH. Each virtual technology has chances to materialize and enter the market. When this happens, virtual organizational structures are turned into real ones. New organizational formats that are generated in the interaction process can be considered as new generations of initial innovation systems, but constructed according to the TH logic. In other words, a fractal tree of innovation systems is forming. National innovation systems, based on the TH model, create the conditions for the appearance of clusters. Cluster structures resemble that of the TH, comprised of parties responsible for normative control, knowledge and wealth generation.

In clusters, these parties no longer represent institutional spheres, but technologies. Informational exchanges between technological trajectories in clusters can lead to the emergence of new technologies. To propel these new technologies into markets, spin-off firms can be organized. The structure of these spin-off firms will also resemble that of a TH, comprising the functions of regulators, knowledge, and wealth generators. The technology, developed by spin-off firms, in turn, can also interact via informational exchange with other technologies, and this may lead to the appearance of other new technologies, supported by new spin-off firms, and so on. The institutional spheres of *G, S,* and *B* can thus act as selection environments and form communicative frameworks that enable cloning of TH structures. This explains the way in which the TH model can account for a ramified structure in different directions.

During the formation of a Triple Helix innovation system inevitably the question of measuring its effectiveness arises [55]. Despite all the importance of effectiveness measurement, currently there are no adequate methods for quantitative evaluation of the processes occurring in the Triple Helix. The task is complicated because, first, the scopes of the actors—Government, University, and Industry—belong to different spheres, characterized by different indices, for which estimations historically were developed using different techniques. The sphere of Government includes changes in institutional structure and legal framework, which can be evaluated in terms of the efficiency of interactions between participants and technology transfer; scientific results are ascertained by the number of publications, patents, and citation index; and industry is evaluated by economic indicators such as income, output, employment, etc.

Second, the analysis of the Triple Helix is hampered by the fact that the principal elements providing dynamics are not the institutional areas of its constituent actors, but the overlapping areas. Existing techniques are focused on the analysis of these institutional areas in their pure form and are not suited for the analysis of overlapping areas. It should also be mentioned that the evaluation of the quality of transactional interactions (the nature and the level of communicational and information-technological exchanges) is just one of the characteristics of the TH. What is more important is the effectiveness of the innovation system functioning, i.e. the quantity and quality of the implemented innovations.

In other words, performance evaluation should take into account the fact that in the knowledge economy the profit maximization criteria is replaced by the criteria of maximizing the economic and technological opportunities, i.e., the system's capacity to generate innovation. From this perspective, the criteria for assessing the effectiveness of an innovation system would consider how effectively new innovations and new markets are developing. The innovativeness depends, among other things, on how branched the structure of an innovation system is.

**7 Summary and Conclusions**

We presented above a qualitative interpretation of the mathematical processes occurring in the Triple Helix model. The model elaborated in the Appendices takes into account the dynamic symmetry in the Triple Helix. As part of a dynamic model of the Triple Helix, we described cyclic processes in the institutional spheres by using wave equation. The wave equation describes the evolution of technological trends. Combined with the internal symmetry of the model, and taking into account local transformations, we derived an additional commutative field, describing the interaction between and among technological trends.

As a result of the previous sections we can draw the following conclusions:

1. The concept and the origin of the non-linearity in innovation systems was clarified. As Etzkowitz and Ranga [19] formulated: "In the communicational framework, University, Industry and Government are seen as coevolving sub-sets of social systems, which are distributed and unstable. They are both *selection environments,* and the institutional communications between them act as *selection mechanisms,* which may generate new innovation environments, ensuring thus the 'regeneration' of the system." We showed that this "regeneration of the system" is due to nonlinear self-interaction of the communication field, originating from the dynamic symmetry of the TH model;

2. In this sense, the Double Helix model can be considered as a linear system, because the institutional communication between DH actors does not act as a selection mechanism that is able to generate new innovation environments, whereas the Triple Helix model possesses a non-linearity at the systems level through which the institutional communication between TH actors can also act as selection mechanisms. Path-dependencies in the order of operation among these selection environments can generate new innovation environments;

3. New innovation environments building on previous ones exhibit a structure reminiscent of a fractal structure. In other words, the applicability of the concept of innovation systems should not be limited to national, regional, sectorial, or technological systems, but can spread far beyond these limits, encompassing even separate local technologies. Each large and small innovation system has a similar TH structures, i.e. comprised of three coevolving actors, that is, regulators, knowledge organizers, and entrepreneurs;

4. The ramified structure of innovation systems is a consequence of the special organizational format formed by three institutional actors who share a functional division of labor among them. When such a structure is created at the national level, it can proliferate clones at all the lower levels of economic activity. This phenomenon was not present in the early industrial economy. The transition from an industrial to a postindustrial phase makes it manifestly visible. In other words, knowledge-based economies can step into the phase of self-organization;

5. We conclude that the system's nonlinearity is also a consequence of waves of innovations, which describe spreading innovation activity in some media, dynamic symmetry in organizational structures, and system's invariants relative to local transformations;

6. The mathematical formulation of the Triple Helix model enables us to specify a more detailed and clear understanding of the processes occurring in the TH. This also allows us to move forward in our understanding of the nature of the processes, and improves our ability to make a prediction. Elaborated mathematical sets of prediction-making techniques consistent with the symmetries of the model can serve as a basis for the quantitative evaluation of economic processes. This can also generate a vast field of follow-up questions for research and ameliorate the functioning of knowledge-intensive markets [56].

## 8 Policy implications

Our model is based upon a well-elaborated quantitative theory in particle physics, and may serve as the base for quantitative economic analysis. These resources were not available in the traditional TH model or other economic models. However, the specification allows us to organize and focus the analysis, has predictive power, explains what has happened, and can form a basis for informing rational actions. The model provides a framework for a flexible system of structuring and interpreting the data, and for comparative analysis.

The transition from an industrial economy to a knowledge-based economy has led to an economic paradigm shift. A market economy may adequately be described in terms of interactions among independent institutional spheres as in the DH model. These kinds of interactions can be expected to lead to technological trajectories that are relatively stable over time. A knowledge-based economy contains an innovation system including the organization and control of knowledge as a systemic component.

An innovation system, based on the TH model, can no longer support stable technological trajectories. Existing trajectories are subject to change, and replaced by subsequent trajectories, as a result of interactions among actors. Newly emerging technologies grow increasingly diversified and their life cycles grow increasingly short. So, adherence to particular technologies may lead to a loss of competitive advantages. Therefore, the core of innovation policies under these new economic conditions should be concentrated not so much on manufacturing, even based on advanced technologies, but on developing new innovation technologies. In other words, one should aim for a shift from the production of material objects to the production of innovation technologies if one wishes to achieve competitive advantages.

**References**

[1] H. Etzkowitz, L. Leydesdorff, The Dynamics of Innovation: from National Systems and "mode 2" to a Triple Helix of university-industry-government relations, Res. Policy, 29 (2) (2000), pp. 109-123.

[2] C. Freeman, The Long Wave Debate, in: T. Vasko (Ed.), Technical Innovations, Diffusion and long Cycles of Economic Development, Springer, Berlin, 1987.

[3] B.-E. Lundvall, Innovation as an interactive process: from user-producer interaction to the national system of innovation. In: Dosi, G., C. Freeman, R. Nelson, G. Silverberg, and L. Soete (Eds.), Technical Change and Economic Theory. London: Pinter, 1988.

[4] B.-E. Lundvall, Introduction, in: B.-E. Lundvall, (Ed.) National systems of innovation: Towards a theory of innovation and interactive learning. London: Pinter, 1992.

[5] R. Nelson, National Innovation Systems: A Comparative Analyses, Oxford University Press, 1993.

[6] M. Porter, The Competitive Advantage of Nations, Free Press, New York, NY, 1990.

[7] M. Porter, On Competition, Harvard Business School, Boston, MA, 1998.

[8] M. Gibbons, C. Limoges, H. Nowotny, S. Schwartzman, P. Scott, M. Trow, The New Production of Knowledge: The Dynamics of Science and Research in Contemporary Societies, SAGE, London, 1994.

[9] L. Leydesdorff, New Models of Technological Change: New Theories for Technology Studies (Epilogue), in: ibid., (1994). pp. 180-192.

[10] H. Etzkowitz, L. Leydesdorff, The Triple Helix – university – industry – government relations: a laboratory for knowledge based economic development, EAAST Review 14 (1995), pp. 14-19.

[11] L. Leydesdorff, The Triple Helix: An evolutionary model of innovations, Res. Policy 29(2) (2000), pp. 243-255.

[12] L. Leydesdorff, G. Zawdie, The Triple Helix Perspective of Innovation Systems, Technol. Anal. Strateg. 22(7) (2010), pp. 789-804.

[13] H. Etzkowitz, M. Ranga, "Spaces": A triple helix governance strategy for regional innovation. In: A. Rickne, S. Laestadius & H. Etzkowitz (Eds.), Innovation Governance in

an Open Economy: Shaping Regional Nodes in a Globalized World , Milton Park, UK: Routledge, 2012.

[14] B. Mandelbrot, The fractal geometry of nature, W.H. Freeman & Co., 1983.

[15] J. A. Shumpeter, Business Cycles: A theoretical, historical and statistical analysis of the Capitalist process, NY, 1939.

[16] E. Peters, Chaos and order in the capital markets, J.Wiley & Sons, 1996.

[17] L. Leydesdorff, M. Meyer, Triple Helix indicators of knowledge-based innovation systems – Introduction to the special issue. Res. Policy 35 (2006), pp. 1441-1449.

[18] L. Leydesdorff, The Knowledge-based economy and the Triple Helix model, Annu. Rev. Inform. Sci. 44 (2010), pp. 367-417.

[19] H. Etzkowitz, M. Ranga, A Triple Helix System for Knowledge-based Regional Development: From 'Spheres" to 'Spaces'. Paper presented at the VIII Triple Helix Conference, Madrid, October 2012.

[20] H. A. Simon, The Organization of Complex Systems. In H. H. Pattee (Ed.), Hierarchy Theory: The Challenge of Complex Systems. New York: George Braziller Inc., 1973.

[21] N. Kondratiev, The Long Waves in Economic Life, Rev. Econ. Statistics, 17 (1935), pp. 105-115.

[22] S. Kuznets, Secular movements in production and prices. Their nature and their Bearing upon Cyclical Fluctuations, Houghton Mifflin, Boston, MA, 1930.

[23] C. Juglar, Des Crises commerciales et leur retour periodique en France, en Angleterre, et aux Etats-Unis,  Guillaumin, Paris, 1892.

[24] J. Kitchin, Cycles and trends in Economic factors. Rev. Econ. Statistics 5 (1923), pp. 10-16.

[25] G. Mensh, Stalemate in Technology – Innovations Overcome the Depression, Ballinger, NY, 1979.

[26] D. Dickson, Technology and Cycles of Boom and Bust, Science 219 (1983), pp. 933-936.

[27] G. Modelsky, W. R. Thompson, Leading Sectors and World Powers: The Coevolution of Global Politics and Economics, Univ. South Carolina Press, Columbia, SC, 1996.

[28] G. Modelsky, Global Political Evolution, Long Cycles and K-Waves, in: T.C. Devezas (Ed.), Kondratieff waves. Warfare and World Security, IOS Press, Amsterdam, 2006.

[29] W. R. Thompson, The Kondratieff Wave as Global Social Process, in:, G. Modelsky, R.A. Denemark (Eds.), World System History, Encyclopedia of Life Support Systems, UNESKO, EOLSS Publishers, Oxford, 2007.

[30] Ch. Papenhausen, Causal Mechanisms of Long Waves, Futures 40 (2008), pp. 788-794.

[31] C. Freeman, C. Perez, Structural crisis of adjustment, business cycles and investment behavior, in: G. Dosi et al. (Eds.) Technical change and economic theory, Pinter, London, 1988.

[32] G. Dosi, Technological paradigms and technological trajectories: a suggested interpretation of the determinants and directions of technical change, Res. Policy, 11 (1982), pp. 147-162.

[33] G. Dosi, The nature of the innovation process, in: G. Dosi et al. (Eds.), Technical change and economic theory, Pinter, London, 1988.

[34] R. Nelson, S. Winter, An Evolutionary theory of Economic Change, Cambridge, Mass.: Harvard University Press, 1982.

[35] W. Dolfsma, L. Leydesdorff, Lock-in and break-out from technological trajectories: Modelling and policy implications, Technol. Forecast. Soc. Change 76 (2009), pp. 932-941.

[36] M. McKelvey, Emerging Environments in Biotechnology, in: H. Etzkowitz , L. Leydesdorff (Eds.), Universities and the Global Knowledge Economy: A Triple Helix of University-Industry-Government Relations, 1997.

[37] J. Van den Ende, W. Dolfsma, Technology push, demand pull and shaping of technological paradigm – Patterns in the development of computer technology, J. Evol. Econ., 15 (2003), pp. 83-99.

[38] L. Leydesdorff, H. Etzkowitz, Emergence of a Triple Helix of University – Industry - Government relations, Sci. Public Policy 23 (1996), pp. 279-286.

[39] R. Nelson, The coevolution of technology, industrial structure and supporting institutions, Ind. Corp.Change, 3 (1994), pp. 47-64.

[40] S. Klepper, Entry, Exit, Growth, and Innovation over the product life cycle, Am. Econ. Rev. 86(3) (1996), pp. 562-583.

[41] F. Malerba, R. Nelson, L. Orsenigo, S. Winter, History friendly models of industry evolution : the case of computer industry, Ind. Corp. Change,  8 (1999), pp. 3-40.

[42] J. Schmookler, Economic Sources of Inventive Activity, J. Econ. Hist. 22 (1962), pp. 1-20.

[43] L. Leydesdorff, D. Rotolo, W. de Nooy, Innovations as a Nonlinear Process, the Scientometric Perspective, and the Specification of an innovation Opportunities Explorer, Technology Analysis and Strategic Management (2013), (in press).

[44] Leydesdorff, L., & Van den Besselaar, P., Evolutionary Economics and Chaos Theory: New Directions in Technology Studies, London and New York, Pinter, 1998.

[45] C. Freeman, L. Soete, The Economics of Industrial Innovation. London: Pinter, 1997.

[46] W. B. Arthur, Competing technologies, increasing returns and lock-in by historical events, Econ. J. 99 (1989), pp.116-131.

[47] E. G. Carayannis, D. F. Campbell, 'Mode 3' and 'Quadruple Helix': toward a 21$^{st}$ century fractal innovation ecosystem, Int. J. Technol. Manage. 46 (2009), pp. 201-234.

[48] E. G. Carayannis, D. F. Campbell, Triple Helix, Quadruple Helix and Quintuple Helix and How Do Knowledge, Innovation, and Environment Relate to Each Other? Int. J. Technol. Manage. 1 (2010), pp. 41-69.

[49] D. Sahal, Technological guideposts and innovation avenues. Res. Policy, 14 (1985), pp. 61-82.

[50] L. Leydesdorff, Ø. Strand, The Swedish System of innovations: Regional Synergies in a Knowledge-Based Economy, Soc. Behav. Sci. 52 (2012), pp. 1-4.

[51] S. Breschi, F. Malerba, Sectoral innovation systems, in: C. Edquist (Ed.), Systems of innovation: Technologies, institutions and organizations, Pinter, London, 1997.

[52] B. Carlsson, R. Stankiewitz, On the Nature, Function and Composition of Technological Systems, J. Evol. Econ. 1 (1991), pp. 93-118.

[53] O. Granstrand, Corporate Innovation Systems. A Comparative Study of Multi-Technology Corporations in Japan, Sweden and the USA, Chalmers University of Technology, 2000.

[54] Yu. Jakubowskiy, B. Karastelev, I. Ivanova, Principles of Construction and Operation of the Innovation System for Project of Development and Production of High-Tech Innovative Wooden House Construction, Pacific Science Review, 14 (2), pp. 194-200, 2012

[55] L. Leydesdorff & I. Ivanova, Mutual Redundancies in Inter-human Communication systems: Steps Towards a Calculus of Processing Meaning, JASIST (in press).

[56] I. Ivanova & L. Leydesdorff, Redundancy Generation in Triple-Helix Relations: Modeled, Measured, and Simulated (in preparation).

[57] H. Etzkowitz, L. Leydesdorff, The endless transition: A "triple helix" of university – industry – government relations, Minerva 36 (1998), pp. 203-208.

[58] McWeeny, R. Symmetry: An Introduction to Group Theory and its Applications. New York: Dover, 2002.

[59] L. Ryder, Quantum field theory. Cambridge Univ. Press, 1986.

[60] C. N. Yang, R. L. Mills, Conservation of Isotopic Spin and Isotopic Gauge Invariance, Phys. Rev, 96 (1954), pp. 191-195.

**Appendix A. Triple Helix model rotational symmetry**

The Triple Helix model can be mapped in a Cartesian coordinate system where three subdynamics, described by actors: *G, S, B*, are spanned orthogonally. Actors are regarded as coevolving, distributed, and unstable social subsystems. Depending on the local environment, inter-institutional interactions function as a selection mechanism that can generate a new innovation environment, which ensures the system regeneration as a new combination of locally distributed modes [10, 57]. This can be described as a rotation of the coordinate system.

$$V' = RV \qquad (A.1)$$

where *R* is 3 × 3 matrix.

These rotations generate a rotation group *O(3)*, i.e. symmetry group that deals with rotations in the three-dimensional space [58]. Any rotation of the vector *V* can be presented as successive rotations around the *G, B, S* axes, defined by three continuous parameters corresponding to the Euler angles $\varphi, \theta, \psi$. The matrices *R* form a non-Abelian Lie group *O(3)*. The non-Abelian feature implies that the order of two successive rotations cannot be interchanged without changing the final result.

From a mathematical perspective, group *O(3)* has a structure similar to the structure of the group *SU(2)*, that consists of 2 × 2 unitary matrices, operating on vectors in two-dimensional complex space $\xi = \begin{pmatrix} \xi_1 \\ \xi_2 \end{pmatrix}$. Vectors *V* and $\xi$ are connected by the following equations:

$$V_S = (\xi_2^2 - \xi_1^2)/2; \quad V_B = (\xi_1^2 + \xi_2^2)/2i; \quad V_G = \xi_1 \xi_2; \qquad (A.2)$$

*i* is the imaginary unit. If the components of *V* are real numbers, the components of the vector $\xi$ are complex numbers. Vector $\xi$ is preferable to use for the sake of calculation convenience.

Using transformations similar to those of (A.2), which relates the vectors *V* and $\xi$, we can pass from the function $c(x,t)$ in Eq. (1) to the function $u(x,t)$. The difference between these two functions is that while $c(x,t)$ is a real-valued function, function $u(x,t)$ is a complex function, which satisfies the wave equation similar to Eq. (5)

$$u_{tt} - a^2 u_{xx} = 0 \qquad (A.3)$$

and has internal symmetry group *SU(2)*. This substitution serves for calculation convenience.

**Appendix B. Dynamic symmetry and gauge fields in Helix-type models**

Utilizing the concept of dynamic symmetry used to describe the strong interactions of elementary particles in quantum field theory, we first consider the processes occurring in the Double Helix model, based on dyadic interaction between Science and Industry, as shown in Fig. B.1.

**Figure B.1**: A balanced model of the *SB* double helix

Eq. (A.3) can be in a customary way obtained from the Euler-Lagrange equation

$$\partial L / \partial u - \frac{\partial}{\partial x_i}\left[\frac{\partial L}{\partial\left(\frac{\partial u}{\partial x_i}\right)}\right] \qquad (B.1)$$

with Lagrangian

$$L = (\partial_i u\, \partial^i u^*)/2 \qquad (B.2)$$

where $x_i = (x, t)$, $\partial_i = \partial/\partial x_i = \partial_t - \partial_x$. Lagrangian (B.2) is invariant under the transformations:

$$u \to e^{-i\Lambda} u;\ u^* \to e^{i\Lambda} u^* \qquad (B.3)$$

where $\Lambda$ - is a real constant. Transformations of the kind (B.3) are called global gauge transformations. The complex field is mathematically equivalent to the field with two real components. If we present $u$ as a vector in a two-dimensional space with components $(u_1, u_2)$, transformation (B.3) would correspond to the rotation of this vector in the plane at an angle of $\Lambda$. Rotations in two dimensions form the group *O(2)*. At the same time, because this transformation is equivalently represented by (A.6), the group in question is the group *U(1)*. The more detailed discussion of Gauge theory can be found in Ryder [59].

In our example, the rotation in a plane corresponds to the dynamic change of two actors' relative contributions. If $\Lambda$ is a constant transformation, (B.3) should be carried out simultaneously for all points in space. This is contrary to the postulate of the finite speed of propagation of the perturbation.

To get around this contradiction, one can assume that $\Lambda$ varies from point to point, i.e., $\Lambda = \Lambda(x_i)$, which accounts for a finite speed of reorganization propagation among participants. This type of transformation is called local gauge transformation.

The condition of invariance under local gauge transformations leads to the appearance of a compensating gauge field. The Lagrangian (B.2) takes the form

$$L = (D_i u D^i u^*)/2 - F_{ij} F^{ij}/4 \qquad (B.4)$$

Here $D_i u$ is covariant derivative defined as:

$$D_i u = (\partial_i + ieA_i)u \qquad (B.5)$$

$F_{ij} = \partial_i A_j - \partial_j A_i$ ; $A_i$ – additional compensatory field, $e$ – constant that determines the strength of the interaction between the field *u* and the field $A_i$. In physics, a field $A_i$ describes electromagnetic field interacting with a matter field *u*, and *e* has the meaning of electric charge which determines the strength of interaction between the fields *u* and $A_i$.

The field equation for $A_i$ follows from the Euler-Lagrange equation, which is identical to (B.1):

$$\partial L / \partial A_i - \partial \left[\frac{\partial L}{\partial(\partial_j A_i)}\right] / \partial x_i \qquad (B.6)$$

Combined equation connecting institutional and communicational frameworks can be written in the form

$$\partial_j F^{ij} = -ie(u^* D^i u - u D^i u^*) \qquad (B.7)$$

We can see that Eq. (B.7) is a linear equation, i.e. it comprises only the first order of the field $A_i$, which describes communication field. And *e* is a measure of the intensity of this interaction.

The above results can be generalized for the case of non-Abelian group *SU(2)*. An example is a field with three components as in Eq. (2). This symmetry leads to non-Abelian gauge fields, which are otherwise known as Yang-Mills fields [60]. The plane rotation is now replaced by a rotation in three-dimensional space of internal symmetry. By analogy with Eq. (B.5) we can write the covariant derivative:

$$D_i u = \partial_i u + g W_i \times u \qquad (B.8)$$

Function $W_i$ is the gauge potential, similar to $A_i$. The analog of $F_{ij}$ is $W_{ij}$:

$$W_{ij} = \partial_i W_j - \partial_j W_i + g W_i \times W_j \qquad (B.9)$$

Last term in Eqs. (B.8) and (B.9) – is a cross product. From the Euler-Lagrange equations with Lagrangian

$$L = (D_i u D^i u^*) - W_{ij} W^{ij} / 4 \qquad (B.10)$$

we can obtain the field equation

$$\partial^j W_{ij} + gW^j \times W_{ij} = g[(\partial_i u) \times u + g(W_{ij} \times u) \times u] \qquad (B.11)$$

which can also be written in the form

$$D^j W_{ij} = g(D_i u) \times u \qquad (B.12)$$

This equation is analogous to the Eq. (B.7). But if the Eq. (B.7) is linear in field $A_i$, i.e. in the absence of the field $u$ it can be rewritten as

$$\partial^i F_{ij} = 0 \qquad (B.13)$$

i.e. it does not contain terms that are the sources of the field, the Eq. (B.12) is nonlinear equation relative to the field $W_i$. In the absence of the field $u$ it takes the form

$$\partial^j W_{ij} = -gW^j \times W_{ij} \qquad (B.14)$$

In other words, the field $W_i$ is by itself the field source and is able to self-generate. This can be expected to lead to a ramified structure of the communication field.